\documentclass[preprint,12pt]{elsarticle}
\usepackage{amsmath}
\usepackage{amssymb}
\usepackage[pdftex,colorlinks]{hyperref}
\usepackage{multirow}

\biboptions{compress}

\begin{document}

\begin{frontmatter}

\title{Symmetric quantum fully homomorphic encryption with perfect security}

\author{Min Liang}\ead{liangmin07@mails.ucas.ac.cn}
\address{Data Communication Science and Technology Research Institute, Beijing 100191, China}

\begin{abstract}
Suppose some data have been encrypted, can you compute with the data without decrypting them? This problem has been studied as homomorphic encryption and blind computing. We consider this problem in the context of quantum information processing, and present the definitions of quantum homomorphic encryption (QHE) and quantum fully homomorphic encryption (QFHE). Then, based on quantum one-time pad (QOTP), we construct a symmetric QFHE scheme, where the evaluate algorithm depends on the secret key. This scheme permits any unitary transformation on any $n$-qubit state that has been encrypted. Compared with classical homomorphic encryption, the QFHE scheme has perfect security. Finally, we also construct a QOTP-based symmetric QHE scheme, where the evaluate algorithm is independent of the secret key.
\end{abstract}

\begin{keyword}
Quantum cryptography \sep homomorphic encryption \sep quantum one-time pad \sep perfect security \sep privacy-preserving quantum computing
\end{keyword}

\end{frontmatter}

\section{Introduction}
Suppose you have encrypted some data, you intend to compute with the data, but you do not want to decrypt them.
Is it possible to compute with the encrypted data without decryption?

This problem is related with the study ``information processing with encrypted data'' or ``privacy-preserving quantum computing''. It has been investigated a lot in modern cryptography, and includes the studies of homomorphic encryption \cite{rivest1978} and blind computing \cite{feigenbaum1986,abadi1989}. It was first considered by Rivest et al. who suggest some homomorphic encryption schemes \cite{rivest1978}. However, these schemes are insecure \cite{brickell1987}. Later, more results are proposed about homomorphic encryption and fully homomorphic encryption \cite{gentry2009,gentry2010}. These are constructed based on some hard computational problems in mathematics, and their security relies on the computational difficulty of these mathematical problems. Thus they have just computational security. In addition, symmetric homomorphic encryption \cite{castelluccia2009,hessler2012,armknecht2011}
is also an useful and interesting research, and it has higher efficiency. Though its encryption key is the same as the decryption key, it is sufficient for many application.

Based on the theory of quantum mechanics, lots of quantum cryptographic protocols \cite{bennett1984,ekert1991,tamaki2003} have been proposed, and have unconditional security. This stimulates us to consider the above problem in the context of quantum information, and intend to get a more secure solution to the privacy-preserving quantum computing. The closely related with the problem includes: 1) blind quantum computing \cite{childs2005,arrighi2006,aharonov2008,broadbent2009,sueki2012,barz2012,vedral2012,morimae2012a,morimae2012b,morimae2010,fitzsimons2012}, where one party delegates quantum computation to another party without revealing his data and result; 2) quantum homomorphic encryption (QHE) \cite{rohde2012}, which allows quantum data to be manipulated without decrypting; 3) quantum private query \cite{giovannetti2008}, which is used to query a database while keeping the query secret; and 4) quantum private comparison \cite{tseng2012,lin2013,guo2013,li2013,liu2013}, which allows two parties to compare their secret data without revealing the secret.

This paper considers the problem of homomorphic encryption in the context of quantum information processing: suppose arbitrary quantum plaintext $\sigma$ has been encrypted (the ciphertext is $\mathcal{E}(\sigma)$), can we perform any quantum operator directly on the ciphertext (without decrypting the ciphertext) and obtain the desired state $\mathcal{E}(T(\sigma))$? The state $\mathcal{E}(T(\sigma))$ is the ciphertext of the result of performing quantum operator $T$ on the original plaintext $\sigma$.

Rohde et al. \cite{rohde2012} studied quantum walk with encrypted data, and proposed a limited QHE scheme using the Boson sampling and multi-walker quantum walk models. This QHE scheme can be used in blind computing of quantum walk. However, QHE has still not been defined and quantum fully homomorphic encryption (QFHE) scheme has not been constructed.

We study the quantum information processing with encrypted data from the aspect of QHE. The definitions of QHE and QFHE have been presented, and some concrete schemes including QFHE scheme have been constructed. All these schemes are constructed based on quantum one-time pad (QOTP)  \cite{boykin2000,ambainis2000}, and have perfect security.

\section{Concepts}
In Ref. \cite{rohde2012}, a limited QHE scheme was presented using the Boson sampling and multi-walker quantum walk models. However, no definition of QHE or QFHE has been given, and no QFHE scheme has been constructed.

QHE can be either symmetric or asymmetric. Both of the two kinds will be defined, but we will focus on the symmetric QHE. All the schemes constructed in the next sections are symmetric QHE schemes.

Quantum states are usually written in the form of density matrices, denoted as $\rho,\sigma$. The output states of quantum operators $T,\mathcal{E}$ are also represented as density matrices.

{\bf Definition 1:} A symmetric quantum homomorphic encryption scheme $\Delta$ has the following four algorithms:
\begin{enumerate}
  \item Key Generating algorithm $KeyGen_\Delta$ is used to generate a key $key$;
  \item $Encrypt_\Delta$ is the encryption algorithm: $\rho=\mathcal{E}(key,\sigma)$, where $\sigma$ is the quantum plaintext;
  \item $Decrypt_\Delta$ is the decryption algorithm: $\sigma=\mathcal{D}(key,\rho)$, where $\rho$ is the quantum ciphertext;
  \item The algorithm $Evaluate_\Delta$ is used to process the quantum ciphertext without decryption. $Evaluate_\Delta$ is associated to a set $\mathcal{F}_\Delta$ of permitted quantum operators. For any quantum operator $T\in\mathcal{F}_\Delta$, according to the key $key$ and quantum ciphertext $\rho$, it performs the algorithm $Evaluate_\Delta(key, T, \rho)$, and can output a quantum state which is just the ciphertext $\mathcal{E}(key,T(\sigma))$.
\end{enumerate}

It should be noticed that there exists a case that the algorithm $Evaluate_\Delta$ is independent of the key. In this case, it should be written as
$Evaluate_\Delta(T,\rho)$ in the above definition. So, the symmetric QHE can be divided into two classes: 1) symmetric QHE with encryption key, where the algorithm $Evaluate_\Delta$ is executed using the encryption key; and 2) symmetric QHE without encryption key, where the algorithm $Evaluate_\Delta$ is executed without using the encryption key.

{\bf Definition 2:} An asymmetric quantum homomorphic encryption scheme $\Delta$ has the following four algorithms:
\begin{enumerate}
  \item Key Generating algorithm $KeyGen_\Delta$ is used to generate two keys -- a public key $pk$ and a secret key $sk$;
  \item $Encrypt_\Delta$ is the encryption algorithm: $\rho=\mathcal{E}(pk,\sigma)$, where $\sigma$ is the quantum plaintext;
  \item $Decrypt_\Delta$ is the decryption algorithm: $\sigma=\mathcal{D}(sk,\rho)$, where $\rho$ is the quantum ciphertext;
  \item The algorithm $Evaluate_\Delta$ is used to process the quantum ciphertext without decryption. $Evaluate_\Delta$ is associated to a set $\mathcal{F}_\Delta$ of permitted quantum operators. For any quantum operator $T\in\mathcal{F}_\Delta$, according to the public key $pk$ and quantum ciphertext $\rho$, it performs the algorithm $Evaluate_\Delta(pk, T, \rho)$, and can output a quantum ciphertext which can be decrypted as $T(\sigma)$ with the secret key $sk$.
\end{enumerate}

Compared with the usual encryption scheme, the QHE scheme has a fourth algorithm $Evaluate_\Delta$, which is used to process the encrypted data. For example, the algorithm $Evaluate_\Delta$ may be described as this: from the key and the quantum operator $T\in\mathcal{F}_\Delta$, it generates another quantum operator $T'$ and performs it on the quantum ciphertext $\mathcal{E}(key,\sigma)$. In this case, the operator $T'$ is related with the operator $T$ and the key, such that
\begin{equation}\label{equ8}
T'(\mathcal{E}(key,\sigma))=\mathcal{E}(key,T(\sigma)).
\end{equation}
The operator $T$ can be regarded as an desired operation on the quantum plaintext. The operation $T'$ corresponding to $T$ is performed on the quantum ciphertext, and can implement the desired operation on the plaintext.

From the definitions of symmetrc/asymmetric QHE scheme, we say the scheme $\Delta$ can handle all the quantum operators in $\mathcal{F}_\Delta$.

The scheme $\Delta$ is (symmetric or asymmetric) quantum fully homomorphic encryption scheme, if it can handle all quantum operators and $Evaluate_\Delta$ is efficient in a similar way as in Ref.\cite{gentry2010}. The quantum operator $T\in\mathcal{F}_\Delta$ may be uncomputable in polynomial time, and suppose its running time is $S_T$. We say $Evaluate_\Delta$ is efficient if there exists a polynomial $g$ such that, for any quantum operator $T\in\mathcal{F}_\Delta$ that can be implemented in time $S_T$, $Evaluate_\Delta$ can be implemented in time at most $S_T\cdot g(\lambda)$, where $\lambda$ is the security parameter.

The security of QHE scheme depends on the security of the encryption algorithm $Encrypt_\Delta$. In the asymmetric QHE scheme, the security also depends on the privacy of the secret key $sk$ while the key $pk$ is public.

\section{Quantum homomorphic encryption}
Some operators are introduced firstly. Four single-qubit operators $X,Y,Z,H$ are shown as follows:
$X=\left(\begin{array}{cc}
  0 & 1\\
  1 & 0
\end{array}\right)$,
$Y=\left(\begin{array}{cc}
  0 & -i\\
  i & 0
\end{array}\right)$,
$Z=\left(\begin{array}{cc}
  1 & 0\\
  0 & -1
\end{array}\right)$,
$H=\frac{1}{\sqrt{2}}\left(\begin{array}{cc}
  1 & 1\\
  1 & -1
\end{array}\right)$.
Two-qubit operator controlled-NOT is denoted as $CNOT$.
The rotation operators about the $\hat{z}$ and $\hat{y}$ axes are defined by $R_z(\theta)=e^{-i\theta Z/2}$ and $R_y(\theta)=e^{-i\theta Y/2}$.
See the Appendix for some of their commutation rules. All the equations used in this paper can be deduced from those commutation rules.

In this section, we will firstly present three QHE schemes on single qubit, whose permitted quantum operators are in the set $\{R_z(\theta)|\theta\in[0,2\pi)\}$, $\{R_y(\theta)|\theta\in[0,2\pi)\}$ or the union of them. Then a QHE scheme for $CNOT$ is also constructed.

In the first QHE scheme, the set of permitted quantum operators is $\{R_z(\theta)|\theta\in[0,2\pi)\}$. The scheme is shown as follows.
\begin{description}
  \item[{\bf QHE scheme 1}]
  \item[$KeyGen_\Delta$:] Randomly select two bits $k\in\{0,1\},j\in\{0,1\}$;
  \item[$Encrypt_\Delta$:] Compute $\rho_c=X^jZ^k\sigma_m Z^kX^j$;
  \item[$Decrypt_\Delta$:] Compute $\sigma_m=Z^kX^j\rho_c X^jZ^k$;
  \item[$Evaluate_\Delta$:] According to $k,j,R_z(\theta)$, it performs quantum operator $R_z(\theta')$ on the given quantum ciphertext $\rho_c$, where $\theta'=(-1)^j\theta$.
\end{description}

It can be verified that
\begin{equation}\label{equ2}
R_z((-1)^j\theta)X^jZ^k=X^jZ^kR_z(\theta),
\end{equation}
where $k\in\{0,1\},j\in\{0,1\}$. According to Eq.(\ref{equ2}), the output state of the algorithm $Evaluate_\Delta$ is
$$R_z((-1)^j\theta)\rho_c R_z((-1)^j\theta)^\dagger=X^jZ^k(R_z(\theta)\sigma_m R_z(\theta)^\dagger)Z^kX^j,$$
which is just the ciphertext of the state $R_z(\theta)\sigma_m R_z(\theta)^\dagger$. Moreover, no decryption is performed during the computing of $Evaluate_\Delta$.
Thus the scheme satisfies the definition of symmetric QHE, and all the unitary transformations in $\{R_z(\theta)|\theta\in[0,2\pi)\}$ are permitted quantum operators of the QHE scheme.

In the second QHE scheme, the set of permitted quantum operators is $\{R_y(\theta)|\theta\in[0,2\pi)\}$. The scheme is shown as follows.
\begin{description}
  \item[{\bf QHE scheme 2}]
  \item[$KeyGen_\Delta$:] Randomly select two bits $k\in\{0,1\},j\in\{0,1\}$;
  \item[$Encrypt_\Delta$:] Compute $\rho_c=X^jZ^k\sigma_m Z^kX^j$;
  \item[$Decrypt_\Delta$:] Compute $\sigma_m=Z^kX^j\rho_c X^jZ^k$;
  \item[$Evaluate_\Delta$:] According to $k,j,R_y(\theta)$, it performs quantum operator $R_y(\theta')$ on the given quantum ciphertext $\rho_c$, where $\theta'=(-1)^{k+j}\theta$.
\end{description}

It can be verified that
\begin{equation}\label{equ3}
R_y((-1)^{k+j}\theta)X^jZ^k=X^jZ^k R_y(\theta),
\end{equation}
where $k\in\{0,1\},j\in\{0,1\}$. According to Eq.(\ref{equ3}), the output state of the algorithm $Evaluate_\Delta$ is
\begin{equation}
R_y((-1)^{k+j}\theta)\rho_c R_y((-1)^{k+j}\theta)^\dagger = X^jZ^k(R_y(\theta)\sigma_m R_y(\theta)^\dagger)Z^kX^j,
\end{equation}
which is just the ciphertext of the state $R_y(\theta)\sigma_m R_y(\theta)^\dagger$. Moreover, no decryption is performed during the computing of $Evaluate_\Delta$.
Thus the scheme also satisfies the definition of symmetric QHE, and all the unitary transformations in $\{R_y(\theta)|\theta\in[0,2\pi)\}$ are permitted quantum operators of the QHE scheme.

{\bf Remark 1:} The relation $R_y((-1)^j\theta)H^jY^k=H^jY^kR_y(\theta)$ can be easily verified. So, in the QHE scheme 2, the encryption/decryption algorithm can also be modified by replacing $X^jZ^k$ with $H^jY^k$, and the algorithm $Evaluate_\Delta$ performs quantum operator $R_y((-1)^j\theta)$ on the given quantum ciphertext.

The third QHE scheme can be constructed by combining the scheme 1 and scheme 2, and its permitted quantum operators are in $\{R_z(\theta),R_y(\theta)|\theta\in[0,2\pi)\}$. The only modification is about the algorithm $Evaluate_\Delta$, and the modified $Evaluate_\Delta$ is described as follow: according to the key $k,j$ and $R_z(\theta)$ (or $R_y(\theta)$), perform quantum operator $R_z((-1)^j\theta)$ (or $R_y((-1)^{k+j}\theta)$) on the given quantum ciphertext $\rho_c$.

Until now, we have presented three symmetric QHE schemes, such that the permitted quantum operators are in $\{R_z(\theta)|\theta\in[0,2\pi)\}$, $\{R_y(\theta)|\theta\in[0,2\pi)\}$, or the union of them. All these schemes are constructed based on QOTP. In these constructions, each permitted quantum operator $T\in\{R_z(\theta),R_y(\theta)|\theta\in[0,2\pi)\}$ is corresponding to another quantum operator $T'$ which is directly performed on the quantum ciphertext. Finding the operator $T'$ is the key point to construct a QHE scheme.

Next, we construct a symmetric QHE scheme that permits $CNOT$ operator. It can be verified that
\begin{equation}\label{equ1}
(X^jZ^k\otimes X^lZ^m)CNOT = CNOT((-1)^{j\cdot m}Z^m\otimes X^j)(X^jZ^k\otimes X^lZ^m),
\end{equation}
where $j,k,l,m$ are arbitrary four bits.
Denote $CNOT'=CNOT((-1)^{j\cdot m}Z^m\otimes X^j)$, then the Eq.(\ref{equ1}) can be written as $CNOT'(X^jZ^k\otimes X^lZ^m)=(X^jZ^k\otimes X^lZ^m)CNOT$, so the operator $CNOT'$ is corresponding to $CNOT$. From this relation, the QHE scheme can be constructed in the same way as follows. In the encryption/decryption algorithm, the QOTP is used to encrypt/decrypt two qubits with $4$-bit key $j,k,l,m$. According to the key $j,k,l,m$, the algorithm $Evaluate_\Delta$ performs $CNOT'$ on the given quantum ciphertext.

\section{Quantum fully homomorphic encryption}
Based on the QHE schemes in the previous section, the QFHE scheme can also be constructed for any $n$-qubit quantum computation (without quantum measurements).

From the Ref. \cite{nielsen2000}, any single-qubit unitary transformation can be written as the form
\begin{equation}\label{equ4}
U(\alpha,\beta,\gamma,\delta)=e^{i\alpha}R_z(\beta)R_y(\gamma)R_z(\delta),
\end{equation}
where $\alpha,\beta,\gamma,\delta$ are real numbers in the $[0,2\pi)$. It means all the single-qubit unitary transformations can be expressed as the set $\{U(\alpha,\beta,\gamma,\delta)|\alpha,\beta,\gamma,\delta\in[0,2\pi)\}$, where $U(\alpha,\beta,\gamma,\delta)=e^{i\alpha}R_z(\beta)R_y(\gamma)R_z(\delta)$.

According to Eq.(\ref{equ2}) and Eq.(\ref{equ3}), it can be concluded that
\begin{eqnarray}\label{equ5}
X^jZ^k U(\alpha,\beta,\gamma,\delta) &=& e^{i\alpha}X^jZ^k R_z(\beta)R_y(\gamma)R_z(\delta),\nonumber\\
&=& e^{i\alpha}R_z((-1)^j\beta)X^jZ^k R_y(\gamma)R_z(\delta),\nonumber\\
&=& e^{i\alpha}R_z((-1)^j\beta)R_y((-1)^{k+j}\gamma)X^jZ^k R_z(\delta),\nonumber\\
&=& e^{i\alpha}R_z((-1)^j\beta)R_y((-1)^{k+j}\gamma)R_z((-1)^j\delta)X^jZ^k ,\nonumber\\
&=& U(\alpha,(-1)^j\beta,(-1)^{k+j}\gamma,(-1)^j\delta) X^jZ^k,
\end{eqnarray}
where $k\in\{0,1\},j\in\{0,1\}$.

From Eq.(\ref{equ5}), any single-qubit unitary operator $U(\alpha,\beta,\gamma,\delta)$ is corresponding to $U(\alpha,(-1)^j\beta,(-1)^{k+j}\gamma,(-1)^j\delta)$. So the QFHE scheme for single-qubit computation is constructed in the same way as the construction of QHE schemes in the previous section.
\begin{description}
  \item[{\bf QFHE on single qubit}]
  \item[$KeyGen_\Delta$:] Randomly select two bits $k\in\{0,1\},j\in\{0,1\}$;
  \item[$Encrypt_\Delta$:] Compute $\rho_c=X^jZ^k\sigma_m Z^kX^j$;
  \item[$Decrypt_\Delta$:] Compute $\sigma_m=Z^kX^j\rho_c X^jZ^k$;
  \item[$Evaluate_\Delta$:] According to $k,j$ and $U(\alpha,\beta,\gamma,\delta)$, it performs the unitary transformation $U(\alpha,(-1)^j\beta,(-1)^{k+j}\gamma,(-1)^j\delta)$ on the given quantum ciphertext $\rho_c$.
\end{description}

According to Eq.(\ref{equ5}), the output state of the algorithm $Evaluate_\Delta$ is
$$X^jZ^k(U(\alpha,\beta,\gamma,\delta)\sigma_m U(\alpha,\beta,\gamma,\delta)^\dagger)Z^kX^j.$$
Thus the scheme also satisfies the definition of symmetric QHE. Moreover, $Evaluate_\Delta$ has the same computational complexity with the quantum operator $U(\alpha,\beta,\gamma,\delta)\in\mathcal{F}_\Delta$, and the set $\{U(\alpha,\beta,\gamma,\delta)|\alpha,\beta,\gamma,\delta\in[0,2\pi)\}$ contains all of the single-qubit unitary transformations, so the above QHE scheme is fully homomorphic for single-qubit computation.

Notice that while intending to perform any given unitary operator on a quantum ciphertext, the unitary operator should be firstly decomposed into the form of Eq.(\ref{equ4}) (or computing the four parameters $\alpha,\beta,\gamma,\delta$).

Next, we present another scheme, which will be shown to be a QFHE scheme, and permits any $n$-qubit unitary transformation.

Any $n$-qubit unitary transformation can be decomposed as the combination of some $CNOT$ and single-qubit unitary transformations. Then any $n$-qubit
computation (without quantum measurements) can be described as a quantum circuit $C$, which consists of some $CNOT$ and single-qubit unitary gates \cite{liang2011}.
From the above constructions of QHE schemes, the $CNOT$ gate or each single-qubit unitary gates $U(\alpha,\beta,\gamma,\delta)$ can correspond to another quantum gate $CNOT'$ or $U(\alpha,(-1)^j\beta,(-1)^{k+j}\gamma,(-1)^j\delta)$. By replacing each gate in quantum circuit $C$ with the corresponding quantum gate, we can obtain a new quantum circuit, denoted as $C'$. The two quantum circuits $C$ and $C'$ satisfy the relation
\begin{equation}\label{equ6}
C'(X^{a_1}Z^{b_1}\otimes\cdots\otimes X^{a_n}Z^{b_n})=(X^{a_1}Z^{b_1}\otimes\cdots\otimes X^{a_n}Z^{b_n})C,
\end{equation}
where $a=a_1\cdots a_n$ and $b=b_1\cdots b_n$ are arbitrary $n$-bit strings selected from $\{0,1\}^n$. The circuit $C'$ is dependent on the circuit $C$ and the two numbers $a,b$.

From Eq.(\ref{equ6}), the QFHE on $n$ qubits can be constructed as follows. QOTP is used as the encryption/decryption algorithms again, and it needs a $2n$-bit key to pefectly encrypt $n$-qubit plaintext, and obtains an $n$-qubit ciphertext. According to the key, the algorithm $Evaluate_\Delta$ replaces each gate in quantum circuit $C$ with its corresponding gate, and gets a new quantum circuit $C'$, and then puts the $n$-qubit ciphertext as the inputs of $C'$. The output state of $C'$ can just be decrypted according to QOTP, and get the result of performing quantum circuit $C$ on the original plaintext. Thus, this QFHE scheme permits any $n$-qubit unitary circuit $C$ to be performed on $n$ qubits that have been encrypted, and no decryption is needed.

Next, the size of quantum circuit $C'$ is analyzed. From the relation $CNOT'=CNOT((-1)^{j\cdot m}Z^m\otimes X^j)$, the gate $CNOT'$ can be implemented by at most $3$ quantum gates (e.g. $CNOT, Z, X$). So, while constructing the new quantum circuit $C'$, each $CNOT$ gate in the circuit $C$ is replaced with at most $3$ quantum gates. On the other hand, each single-qubit gate in $C'$ is only corresponding to $1$ single-qubit gate. Then, we can conclude that, the size of the new circuit $C'$ is not more than three times of the size of $C$. Thus, the complexity of $Evaluate_\Delta$ is at most three times than that of the circuit $C$, and then the algorithm $Evaluate_\Delta$ is efficient.

It follows from the above analysis that, this scheme is indeed a QFHE scheme, which permits any $n$-qubit unitary transformation.

\section{Analysis}
The security of these QHE and QFHE schemes is analyzed.

{\bf Theorem 1:}
All the QHE and QFHE schemes proposed in the previous sections are perfectly secure.

{\bf Proof:}
Firstly, it can be verified that \cite{boykin2000}
\begin{equation}\label{equ7}\frac{1}{2^{2n}}\sum_{a,b\in\{0,1\}^n}X^aZ^b\tau Z^b X^a=\frac{1}{2^n}I_{2^n},\end{equation}
where $\tau$ is arbitrary $n$-qubit state.

In all these schemes, QOTP is used as the encryption/decryption algorithm. Suppose the single-qubit plaintext $\sigma_m$ is encrypted by QOTP using secret key $j,k\in\{0,1\}$, which is unknown to the attacker. With regard to the attacker, the output of the algorithm $Encrypt_\Delta$ is $\frac{1}{2^2}\sum_{j,k\in\{0,1\}}X^jZ^k\sigma_m Z^kX^j$, which is just the totally mixed state according to Eq.(\ref{equ7}). Moreover, the output of $Evaluate_\Delta$ with regard to the attacker is
\begin{eqnarray}
&& \frac{1}{2^2}\sum_{j,k\in\{0,1\}}U(\alpha,(-1)^j\beta,(-1)^{k+j}\gamma,(-1)^j\delta)\rho_c U(\alpha,(-1)^j\beta,(-1)^{k+j}\gamma,(-1)^j\delta)^\dagger \nonumber\\
&&= \frac{1}{2^2}\sum_{j,k\in\{0,1\}}X^jZ^kU(\alpha,\beta,\gamma,\delta)\sigma_m U(\alpha,\beta,\gamma,\delta)^\dagger Z^kX^j =\frac{1}{2}I_{2}, \nonumber
\end{eqnarray}
which is also the totally mixed state. Similarly, the same result can be obtained for the $n$-qubit QFHE scheme. Thus, with regard to the attacker, the outputs of the algorithms $Encrypt_\Delta$ and $Evaluate_\Delta$ in each QHE/QFHE scheme are totally mixed states.
It means these QHE and QFHE schemes are also perfectly secure, and the attacker can obtain nothing about the plaintext and the result of evaluation.
$\hfill{~}\Box$

All the QHE schemes presented here are constructed based on QOTP. It should be noticed that, the algorithm $Evaluate_\Delta$ is dependent on the secret key.
This may restrict the application of these schemes, because only the owner of the secret key can process the encrypted qubits correctly. Does there exist a QOTP-based symmetric QHE scheme, in which the algorithm $Evaluate_\Delta$ is independent of the secret key? Next, we can present a positive answer to this question.

For convenient, suppose $C$ is a permitted unitary operator in the QHE scheme, and $C'$ is the unitary operator which corresponds to $C$. The algorithm $Evaluate_\Delta$ performs $C'$ on the ciphertext $X^jZ^k|\varphi\rangle$, and can get the result $X^jZ^kC|\varphi\rangle$.

{\bf Lemma 1:}
In the QOTP-based QHE scheme, if the operator $C'$ is independent of the secret key, and satisfies the condition $(X^jZ^k)C=C'(X^jZ^k),\forall j,k$, then $C\in\{e^{i\theta}I|\theta\in[0,2\pi)\}$.

{\bf Proof:}
From the above condition, we know $(X^jZ^k)C(Z^kX^j)$ is independent of the secret key $j,k$. Because $\{X^jZ^k|k,j\in\{0,1\}^n\}$ is a complete orthogonal basis in the $n$-qubit Hilbert space, any $n$-qubit unitary operator $C$ can be represented as
$C=\sum_{a,b\in\{0,1\}^n}\alpha_{a,b}X^aZ^b$. Then $$X^jZ^kCZ^kX^j=\sum_{a,b\in\{0,1\}^n}\alpha_{a,b}(-1)^{a\cdot k\oplus b\cdot j}X^aZ^b,$$ where ``$\cdot$'' denotes inner product modular 2. Because $X^jZ^kCZ^kX^j$ is independent of $k,j$, it can be deduced that, $\forall (k_1,j_1)\neq(k_2,j_2)$, $$\sum_{a,b}\alpha_{a,b}\left((-1)^{a\cdot k_1\oplus b\cdot j_1}-(-1)^{a\cdot k_2\oplus b\cdot j_2}\right)X^aZ^b=0.$$
So, $$\alpha_{a,b}\left((-1)^{a\cdot k_1\oplus b\cdot j_1}-(-1)^{a\cdot k_2\oplus b\cdot j_2}\right)=0,\forall a,b,\forall (k_1,j_1)\neq(k_2,j_2).$$ Because it is impossible that $\alpha_{a,b}=0,\forall a,b$ (otherwise $C=0$), we have $$a\cdot k_1\oplus b\cdot j_1=a\cdot k_2\oplus b\cdot j_2, \exists a,b,\forall (k_1,j_1)\neq(k_2,j_2).$$
Then $$(a,b)\cdot (k_1-k_2,j_1-j_2)=0, \exists a,b,\forall (k_1-k_2,j_1-j_2)\neq (0,0).$$
Thus it concludes that $(a,b)\equiv (0,0)$, and $C=\alpha_{0,0}X^0Z^0=e^{i\theta}I_{2^n}$, $\forall\theta\in[0,2\pi)$. Thus complete the proof.
$\hfill{~}\Box$

Because the global phase in a quantum state does not influence the result of measurement (the global phase has no influence on the density matrix), the condition in Lemma 1 can be relaxed as this: $e^{i\beta}(X^jZ^k)C=C'(X^jZ^k),\forall j,k$, where $\beta$ depends on $j,k$.

{\bf Lemma 2:}
In the QOTP-based QHE scheme, if the operator $C'$ is independent of the secret key, and satisfies the condition $e^{i\beta}(X^jZ^k)C=C'(X^jZ^k),\forall j,k$, then all the operators in $\{e^{i\theta}X^aZ^b|a,b\in\{0,1\}^n,\theta\in[0,2\pi)\}$ are permitted unitary operators.

{\bf Proof:}
It can be verified that
$$(-1)^{a\cdot k\oplus b\cdot j}(X^jZ^k)(e^{i\theta}X^aZ^b)=(e^{i\theta}X^aZ^b)(X^jZ^k),\forall j,k.$$	
Let $C=C'=e^{i\theta}X^aZ^b$, which is independent of $j,k$. Thus complete the proof.
$\hfill{~}\Box$

From the two lemmas, we can conclude the following result.

{\bf Theorem 2:}
If the algorithm $Evaluate_\Delta$ is independent of the secret key, then the QOTP-based symmetric QHE scheme permits any of the unitary operators in the set $\{e^{i\theta}X^aZ^b|a,b\in\{0,1\}^n,\theta\in[0,2\pi)\}$.$\hfill{~}\Box$

Now we have shown there exists a symmetric QHE scheme, which satisfies the condition: $Evaluate_\Delta$ is independent of the secret key. However, it is obvious that it cannot be fully homomorphic.

There is another point that is worth to notice. It can be concluded from Eq.(\ref{equ8}) that, given any $T$, if the encryption algorithm $\mathcal{E}$ is replaced by another encryption algorithm $\mathcal{E}'$, then the operation $T'$ will changes accordingly. The relationship between $T'$ and $T$ depends on the encryption algorithm $\mathcal{E}$. In all our schemes,
QOTP is chosen as the encryption algorithm, and this makes for our simple constructions.

\section{Discussions}
The symmetric QFHE scheme has been constructed. Suppose you have encrypted a qubit with this scheme, you can perform any single-qubit unitary operator on the qubit without decryption. If you intend to delegate the unitary operator to another party, you have to preshare the key with that party, thus that party may decrypt it and obtain the original qubit. So the symmetric QFHE scheme proposed here cannot be used in blind computing. However, you can still delegate the computation to the trusted party by using the QFHE scheme, which can prevent the malicious parties from obtaining the data and the result of computation.

The symmetric QFHE scheme may be used in secure multiparty quantum computation \cite{dupuis2010,dupuis2012}, where $n$ parties participate in a computation (e.g. an evaluation of a unitary circuit $C$). Suppose the circuit $C$ is given $n$ quantum inputs and outputs $n$ quantum states, each of which is desired by one party. A naive multiparty computation protocol is given as follows. Each participant preshares a secret key with trusted third party (TTP) through quantum key distribution (QKD), and encrypts his inputs with his secret key according to the QFHE scheme, and then sends his ciphertext to TTP. TTP performs quantum circuit $C'$ (which is relative to $C$ and all the keys shared between TTP and every party) on all the ciphertexts, and sends back the results to each party. Then each party can decrypt the received state with his secret key, and obtain the desired result.

Besides the homomorphic encryption, blind computing is another kind of study about privacy-preserving computation. Actually, homomorphic encryption and blind computing are two closely related research directions. Homomorphic encryption scheme can be used in blind computing in the following two cases: (1) Homomorphic encryption scheme is symmetric, and the algorithm $Evaluate_\Delta$ is independent of the key $key$; (2) Homomorphic encryption scheme is asymmetric. In blind computing, the decryption $f(c)\rightarrow f(m)$ is unnecessarily  the inverse of the encryption $m\rightarrow c$ (See Refs.\cite{feigenbaum1986,childs2005}). However, in the homomorphic encryption scheme, the decryption algorithm is just the inverse of the encryption algorithm.

In the QFHE scheme, $2n$ bits of key are needed to encrypt $n$ qubits with perfect security. To reduce the length of key, the QOTP can be replaced with approximate QOTP \cite{ambainis2004}. However, the security would be slightly weaker.

This paper considers the symmetric QHE, and the asymmetric QHE is only defined but not constructed. So the open problem is: how to construct an asymmetric QHE scheme, where the algorithm $Evaluate_\Delta$ depends only on the public key $pk$ but not the secret key $sk$? Or you can consider how to modify the quantum public-key encryption scheme in Ref.\cite{liang2012}, such that it becomes an asymmetric QHE scheme.
If this goal were achieved, the computing on the quantum ciphertext could be securely outsourced, and then the blind quantum computing would be implemented in this way.

\section{Conclusions}
This paper defines symmetric and asymmetric QHE, and proposes four symmetric QHE schemes that permit all the quantum operators in the set $\{R_z(\theta)|\theta\in[0,2\pi)\}$, $\{R_y(\theta)|\theta\in[0,2\pi)\}$ or $\{CNOT\}$. Moreover, we construct a symmetric QFHE scheme, which permits any unitary operator on single qubit. Then the QFHE scheme is extended to permit any $n$-qubit unitary operator. All these schemes are constructed based on QOTP, and have perfect security. In the QFHE scheme, the algorithm $Evaluate_\Delta$ depends on the secret key. Finally, we proposed a symmetric QHE scheme, in which $Evaluate_\Delta$ is independent of the secret key.

\section*{Appendix}
This section lists some commutation rules about the operators $X,Y,Z,H$, $CNOT,R_z(\theta),R_y(\theta)$. Here, $j,k,l,m$ are single-bit numbers.
\begin{eqnarray}
&& Z^kX^j=(-1)^{j\cdot k}X^jZ^k, \nonumber\\
&& R_z(\theta)X^j=X^jR_z((-1)^j\theta),\forall\theta\in[0,2\pi), \nonumber\\
&& R_y(\theta)X^j=X^jR_y((-1)^j\theta),\forall\theta\in[0,2\pi), \nonumber\\
&& R_y(\theta)Z^k=Z^kR_y((-1)^k\theta),\forall\theta\in[0,2\pi), \nonumber\\
&& R_y(\theta)H^j=H^jR_y((-1)^j\theta),\forall\theta\in[0,2\pi), \nonumber\\
&& CNOT(X^j \otimes I)=(X^j \otimes X^j)CNOT, \nonumber\\
&& CNOT(Z^k \otimes I)=(Z^k \otimes I)CNOT, \nonumber\\
&& CNOT(I \otimes X^l)=(I \otimes X^l)CNOT, \nonumber\\
&& CNOT(I \otimes Z^m)=(Z^m \otimes Z^m)CNOT. \nonumber
\end{eqnarray}

\end{document}